\documentclass{iau307}
\usepackage{graphicx}
\usepackage{natbib}
\usepackage{url}
\usepackage{dtklogos}
\bibpunct{(}{)}{;}{a}{}{,}

\newcommand{\ioni}[2]{{#1\,\sc{#2}}}
\newcommand{\vs}{$v$~sin~$i$}
\newcommand{\vm}{$v_{\rm mac}$}
\newcommand{\Teff}{T$_{\rm eff}$}
\newcommand{\grav}{log~$g$}
\newcommand{\kms}{km~s$^{-1}$}

\title[The macroturbulent broadening - pulsation connection] 
{Asteroseismology of OB stars with hundreds of single snapshot 
spectra \\ {\Large (and a few time-series of selected targets)}}

\author[S. Sim\'on-D\'iaz]   
{S. Sim\'on-D\'iaz$^{1,2}$
 }

\affiliation{$^1$Instituto de Astrof\'isica de Canarias, 38200 La Laguna, Tenerife, Spain\\ 
email: {\tt ssimon@iac.es} \\[\affilskip]
$^2$Departamento de Astrof\'isica, Universidad de La Laguna, 38205 La Laguna, Tenerife, Spain
}

\pubyear{2014}
\volume{307} 
\pagerange{}
\jname{New windows on massive stars: asteroseismology, interferometry, and spectropolarimetry}
\editors{G. Meynet, C. Georgy, J.H. Groh \& Ph. Stee, eds.}

\begin{document}

\maketitle

\begin{abstract}

Imagine we could do asteroseismology of large samples of OB-type stars by using just 
one spectrum per target. That would be great! But this is probably a crazy and stupid idea. 
Or maybe not. Maybe we have the possibility to open a new window to investigate stellar 
oscillations in massive stars that has been in front of us for many years, but has not 
attracted very much our attention: the characterization and understanding of the so-called 
macroturbulent broadening in OB-type stars.
\keywords{stars: early-type, 
stars: oscillations (including pulsations), 
line: profiles, 
techniques: spectroscopic}
\end{abstract}

\firstsection 
\section{Introduction\label{Sec1}}
\subsection{From macroturbulent broadening to pulsations (a bit of context)}
First references to macroturbulent broadening in O-type stars and early-B
supergiants can be found in \cite{Str52}, \cite{Sle56}, \cite{Con77}, \cite{Pen96}, and \cite{How97}.
Based on the lack of this type of stars with sharp absorption lines, these authors 
proposed that rotation was not the only broadening mechanism shaping their line-profiles.
This hypothesis was definitely confirmed with the advent of high-quality spectroscopic observations
and its analysis by means of adequate techniques \citep[e.g.,][]{Rya02, Sim07, Sim14}.

Modern measurements of the non-rotational broadening component soon allowed to discard that it 
was produced by any type of large scale turbulent motion \citep{Sim10}. In this context, 
\cite{Aer09} revived an alternative scenario to explain its physical origin: the {\em pulsational 
hypothesis}.
\subsection{The pulsational view of macroturbulent broadening (and vice versa)}
The basic idea proposed by \citeauthor{Aer09} (also previously indicated by \citeauthor{Luc76}
\citeyear{Luc76}, and \citeauthor{How04} \citeyear{How04})
is that the collective pulsational velocity broadening due to gravity modes could be a viable 
physical explanation for macroturbulence in massive stars. The presence of an important (variable) 
pulsational broadening component is firmly established in B dwarfs and giants located in
the $\beta$-Cep and SPB instability domains \cite[e.g.,][and references therein]{Aer14}; however, the macroturbulent-pulsational broadening
connection in O-type stars and B supergiants (B~Sgs), located in a region of the HR diagram which to-date 
is (by far) less explored and understood from an asteroseismic point of view, required further 
(observational) confirmation.

As discussed by \cite{Sim12}, this hypothesis might also open the possibility to use
macroturbulent broadening as a spectroscopic indicator of the occurrence 
of a certain type of stellar oscillations in massive stars. Obviously, this alternative 
strategy will never be able to compete with a detailed asteroseismic study; however, if 
at some point the macroturbulent\,--\,pulsational broadening connection is confirmed, 
this spectroscopic feature could become a cheap, single-snapshot way to detect and 
investigate pulsations in massive stars from a complementary perspective.
{\em Reward may be juicy, worth a shot!}
\subsection{The macro-pulsa project}
Motivated by the characterization of the macroturbulent broadening in the whole O and B-type 
star domain, and the investigation of its postulated pulsational origin, we have compiled during 
the last 6 years a large high-resolution spectroscopic dataset comprising more than 3500~spectra 
of about 500 Galactic O4-B9 stars (including all luminosity classes). 
In this talk I highlight the most important results obtained to-date in this context from the 
exploitation of this unique spectroscopic dataset.
\section{Line-broadening in OB stars: a single-snapshot overview\label{Sec2}}
\subsection{Observations, methods, and some results}
The spectroscopic observations are drawn from the {\em IACOB spectroscopic database of Northern
Galactic OB star}. Last described in \citet[][SDH14]{Sim14}, the database now also include 
new\footnote{The {\em IACOB database} initially comprised FIES@NOT spectra 
of bright O- and early B-type stars. Given the similar
capabilities and performance of HERMES@MERCATOR, we have decided to join together the IACOB
spectra with those gathered as part of the IACOB-sweG (P.I. Negueruela) and macro-pulsa projects
(P.I. Sim\'on-D\'iaz), mainly using HERMES.}
HERMES spectra of bright B2\,--\,B9 stars. We have discarded from the
initial sample all stars detected as SBx (x$\ge$2) and applied the methodology described in SDH14 to 
disentangle the rotational (\vs) and macroturbulent (\vm, using a radial-tangential prescription) 
broadening components from the \ioni{O}{iii}$\lambda$5592, \ioni{Si}{iii}$\lambda$4552, 
\ioni{Mg}{ii}$\lambda$4481 or \ioni{C}{ii}$\lambda$4267 line-profiles. 
A first estimation of the effective temperature (\Teff) and gravity (\grav) 
in the O- and B-star samples was obtained by means of updated versions of the grid-based automatized
tools described in \cite{Sim11} and \cite{Cas12}, respectively.

\begin{figure}[h]
 \centering
  \includegraphics[width=10.cm,angle=0]{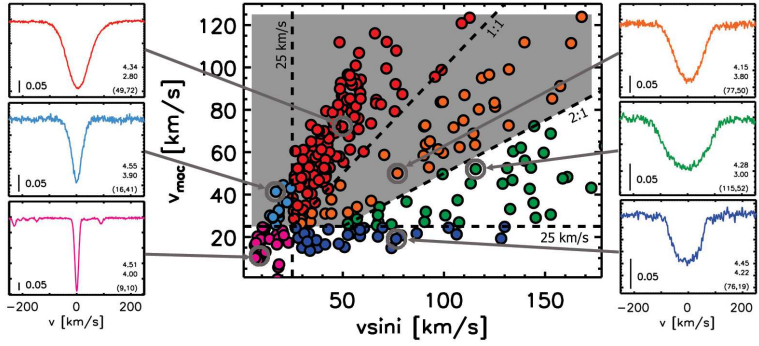}
 \caption{\vs\,--\,\vm\ diagram including 100 O- and 200 B-type stars. The diagram is 
 divided into six regions depending of the relative contribution of the rotational and
 macroturbulent broadenings to the global shape of the line-profiles (see also notes in SDH14). 
 The three regions with a clear contribution of the macroturbulent broadening are shadowed in
 gray. Panels to the left and right show one illustrative profile per zone. Numbers quoted
 in each panel are the (normalized) flux scale, log~\Teff, \grav, and the measured \vs\ and 
 \vm\ (both in \kms). 
}
 \label{fig1}
\end{figure}

Fig.~\ref{fig1} summarizes the results of the line-broadening analysis of our final 
sample in the \vs\,--\,\vm\ diagram. 
Stars with \vs$\ge$180~\kms\ are excluded from this discussion (see SDH14).
We differentiate six regions in the diagram depending of the 
relative contribution of the rotational and macroturbulent broadenings to the 
line-profile. The most important result are 
(1) the strong correlation found in stars with \vm$\,\ge\,$\vs, and (2) that stars with 
similar (\vs,\vm) combinations present almost identical profiles independently of their 
spectral type and luminosity class.

\begin{figure}[h]
 \centering
  \includegraphics[width=5.cm,angle=90]{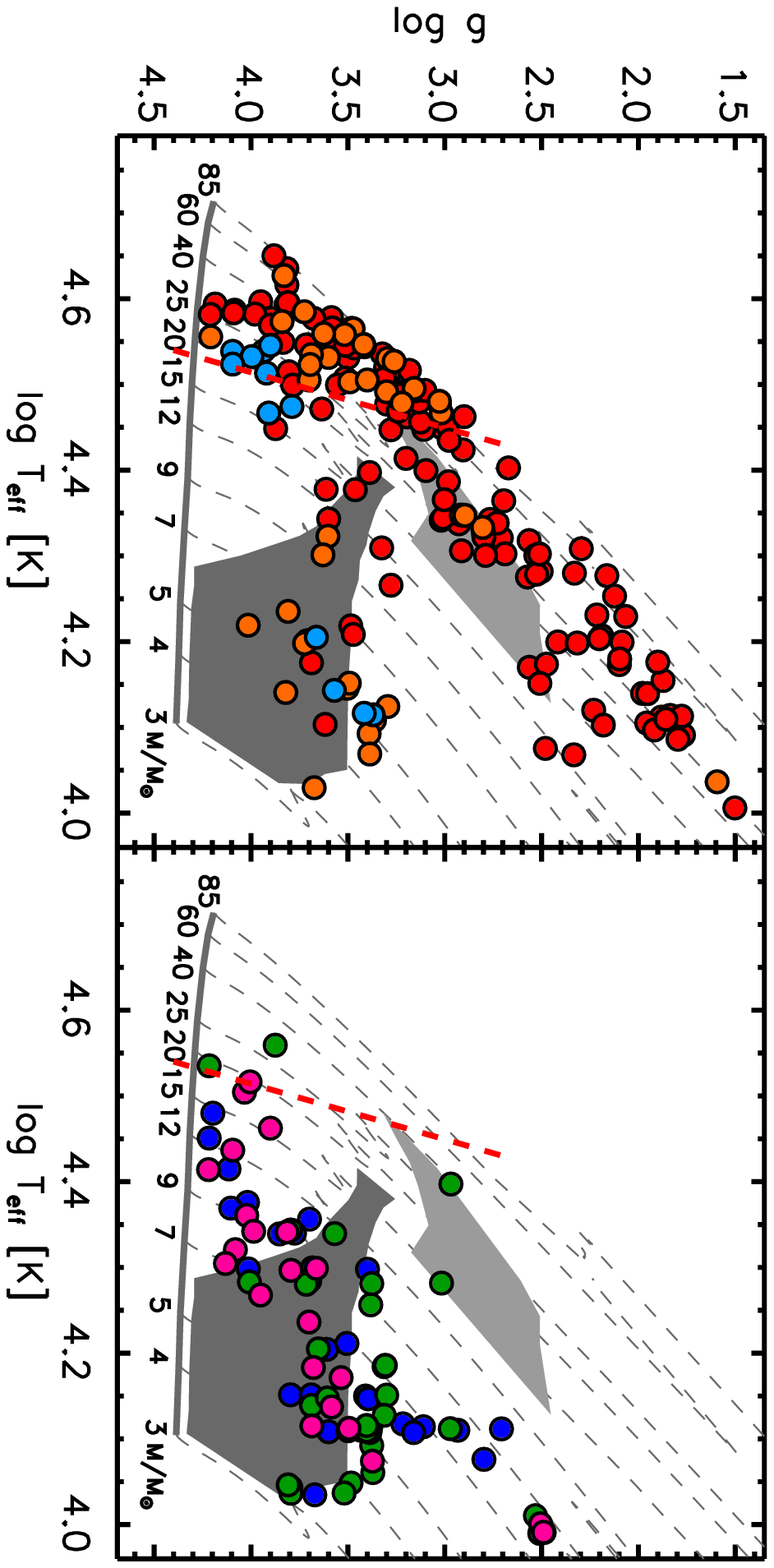}
 \caption{log~\Teff\,--\,\grav\ diagrams with the analyzed stars, separated by cases with (right)
 and without (left) a clear macroturbulent broadening component. The red tilted line separates O- 
 and B-stars. In the background: evolutionary tracks from \cite{Eks12} and high-order g-mode 
 instability strips from \citet[][OPAL, GN93, MS, M$\le$18 M$_{\odot}$; dark gray]{Mig07} and 
 \citep[][OPAL, GN93, Vink mass-loss rates, MS and post-MS, M$\ge$10 M$_{\odot}$; light gray]{God11}.
 }
 \label{fig2}
\end{figure}

Stars with and without a clear macroturbulent broadening component (the former grouped in the shadowed
gray region in Fig.~\ref{fig1}) are located separately in the log~\Teff\,--\,\grav\ diagram in 
Fig.~\ref{fig2}. The two diagrams are complemented with state-of-the-art evolutionary tracks 
and computations of high-order g-mode instability domains. Interestingly, there is a clear separation
between the regions in the HR diagram where the stars with/without an important contribution
of the macroturbulent broadening are located. However, there does not seem to be any clear correlation
between these regions and the predicted instability domains for high order g-modes. Especially challenging 
for the pulsational hypothesis (in terms of g-modes) are the O-stars and the late-B Sgs.

\newpage 

\section{Macroturbulent broadening and line-profile variability\label{Sec3}}

\subsection{Observations, methods, and some results}

During the last 6 years we have been compiling spectroscopic time-series (with
 both FIES and HERMES) of a selected a subsample of 10 O stars and 11 B~Sgs. All 
 targets (except three: one O-dwarf, one B-dwarf, and one late-B~Sg) were selected as having a 
dominant macroturbulent broadening component. For a few of them we already have more than 150 spectra, 
for the rest we count on a few dozens of spectra. The main bulk of the observations are 
characterized by a cadence of 4-8 spectra per night separated by 0.5-1.5~h. The typical exposure time is 
between a few seconds and 15 min. The length of each separated run is typically a few nights (being 10 nights 
the longest one). We are investigating line-profile variability (LPV) using the moment method \citep{Aer92}, and 
line-broadening variability using the {\sc iacob-broad} tool (SDH14).

\begin{figure}[t!]
 \centering
 \includegraphics[width=3.7cm,angle=90]{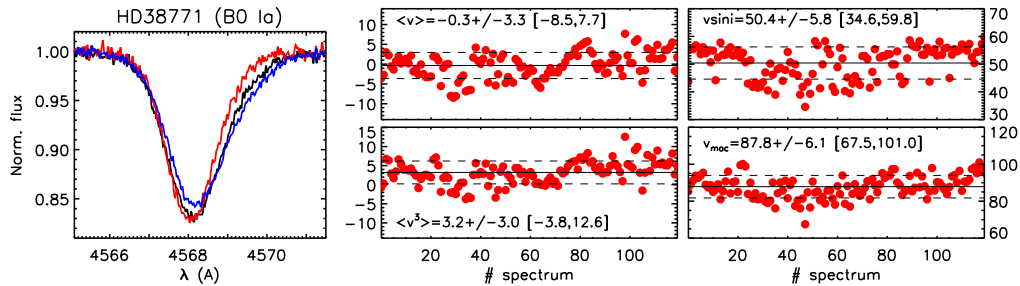}
 \caption{Illustrative example of the type of line-profile variability found in the
 sample of O stars and B supergiants for which we have obtained spectroscopic time-series. 
 [Right panels] Variability in the first and third moments of
 the line-profile (center), and the two parameters defining the line-broadening (right).
 Quoted numbers indicate the mean value, standard deviation, minimum and maximum values, respectively.
 All quantities in \kms\ except for $\langle v^3 \rangle$ (in 10$^4$~km$^3~$s$^{-3}$).
 [Left panel] Three characteristic profiles having $\langle v^3 \rangle$=0 (black) and maximum negative/positive
 skewness (red and blue, respectively).}
 \label{fig3}
\end{figure}

\begin{minipage}{0.40\textwidth}
 \centering
 \includegraphics[width=4.2cm,angle=90]{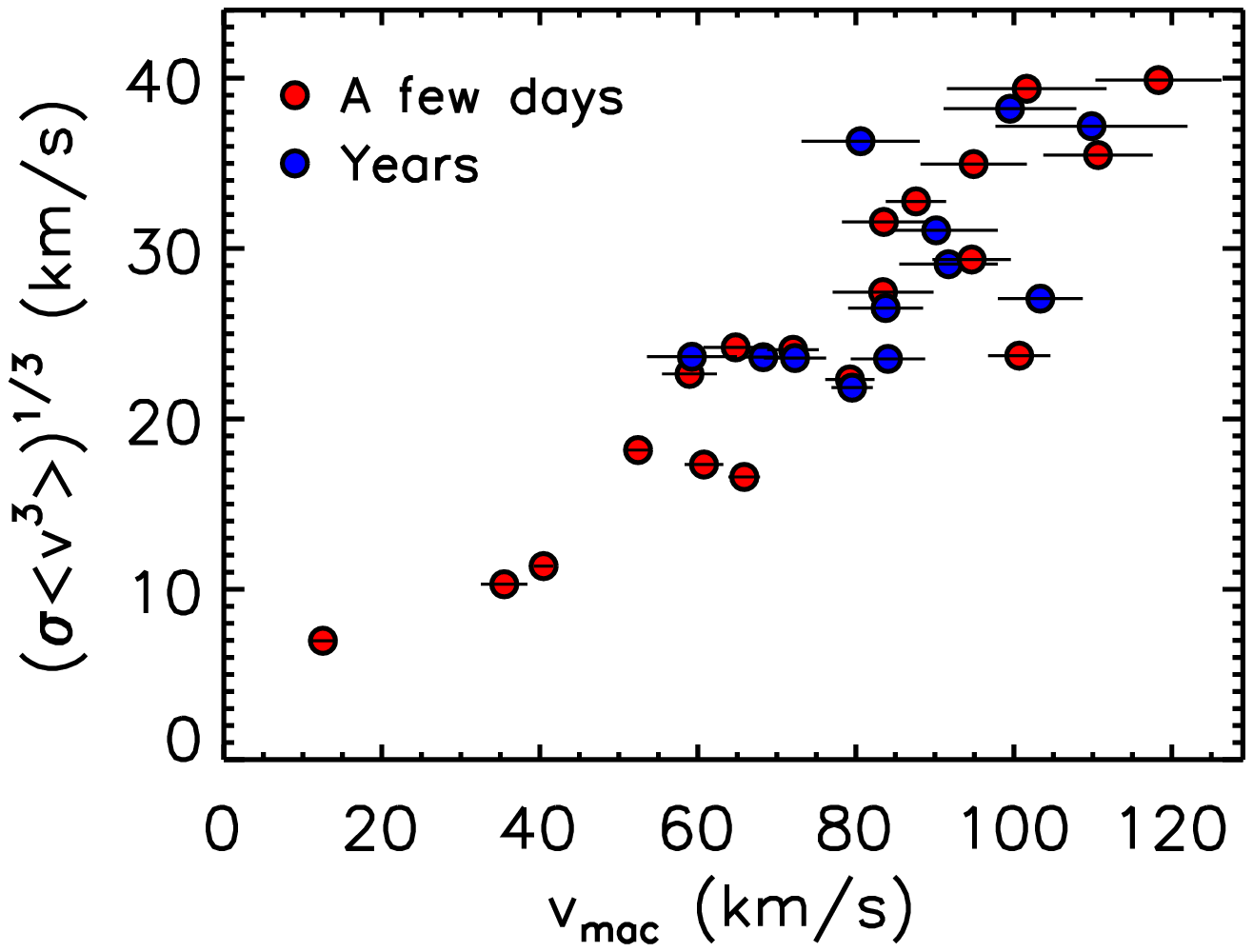}
\end{minipage}
\begin{minipage}{0.05\textwidth}
.
\end{minipage}
\begin{minipage}{0.47\textwidth}
{\small {\bf Figure 4.} Observational correlation found between the size of the macroturbulent
broadening and the amplitude of variation of the skewness of the line-profile. 
Horizontal lines indicate the range of variability of \vm. Close-by red and blue symbols correspond
to the same stars but quantities computed covering different temporal scales (a few days and
years, respectively).} 
\vspace{0.5cm}
\end{minipage}


The frequency analysis of our spectroscopic time-series (prior to any mode identification) is a 
complicated task due to the non-optimal observational cadence and the complexity of the
line-profile variability detected in this type of stars (C. Aerts, these proceedings).
However, we can already highlight some first observational results which must be taken into 
account by any attempt to explain the physical origin of the macroturbulent broadening: 
\begin{itemize}
\item Line-profiles of all the studied stars show LPVs (with similar characteristics as
those illustrated in Fig.~\ref{fig3}). Despite the variability, the global characteristic 
{\em V-shape} of the profiles remains {\em roughly} constant. Indeed, a similar shape can be
found in very short ($\sim$~a few seconds) and much longer (e.g. co-adding all the spectra
obtained during 6 years) exposures.
\item The typical peak-to-peak amplitude of variability of the first moment ($\langle v \rangle$, centroid) 
is $\sim$ a few \kms. We also find small variability ($\le$10\%) in the \vs\ and \vm\ measurements 
provided by {\sc iacob-broad}.
\item We find strong correlations between \vm\ and the amplitude of variability of
the third moment ($\langle v^3 \rangle$, skewness); this was shown for the first time in \cite{Sim10} and is now
confirmed with a larger sample and a longer time-span (see Fig.~4).
\item Although the precise identification of the frequencies of line-profile variability is difficult, 
we find some systematic hits of all stars showing multi-periodic variability in the high-order 
g-mode frequency domain ($\sim$ a few hours to several days). However, the possibility of 
non-strictly-periodic variability cannot be discarded.
\end{itemize}

\section{Concluding remarks\label{Sec4}}

The presence of macroturbulent broadening in the whole O-type star domain, where instability domain 
computations predict no excitation of high-order gravity mode, seems to be a strong empirical 
challenge to the pulsational hypothesis (at least in reference to this type of stellar oscillations). 
The same occurs for the late-B Sgs. However, the detected line-profile variability in all studied
stars with macroturbulent broadening, and the empirical correlations found between \vm\ and
$\sigma$($\langle v^3 \rangle$) points towards to existence of a connection between the physical 
origin of this spectroscopic feature and stellar variability phenomena occurring in the 
photosphere of stars with masses above 15~M$_{\odot}$.

The understanding of the macroturbulent broadening is still an open issue, but the new 
observational material and the strategy proposed here will certainly
help us to find the solution of a long-standing question in the field of massive stars. Even 
if the pulsational hypothesis may be rejected at some point, any other proposed scenario must fulfill the
empirical (single snapshot and time-dependent) constraints described in Sects.~\ref{Sec2} and \ref{Sec3}. 
In this context, it is important to have also present other related studies proving complementary 
empirical constraints \citep[e.g.,][]{Sun13a,Mar14}, or alternative scenarios to the origin of the macroturbulent
broadening \cite[e.g.,][]{Can09}, and maintain the initiated synergies between this research 
line and Asteroseismology. A new window on massive stars is now fully open.\\

\textbf{Acknowledgements} 

{\small Although they are not included in the co-author list of this proceeding, I would like
to thank many colleagues for interesting 
discussions during the development of this project. In particular to A.~Herrero, C.~Aerts, 
P.~Degroote, J.~Puls, M. Godart, N.~Markova, and E.~Moravveji. Special thanks to
N.~Castro for providing me with the stellar parameters of the B star sample, and to a
long list of observing colleagues who are making the compilation of the observational 
data needed for this project more manageable: I. Negueruela, R.~Dorda, I.~Camacho, K.~R\"ubke, P.~de~Cat, 
S.~Triana, C.~Gonz\'alez, A. Gonz\'alez, E.~Niemczura, D. Drobek. Last, but not least, the 
NOT and MERCATOR staff for their high competence and always useful assistance. This work 
has been funded by the Spanish Ministry of Economy and Competitiveness 
under the grants AYA2010-21697-C05-04, and Severo Ochoa SEV-2011-0187, and by 
the Canary Islands Government under grant PID2010119.}

\bibliographystyle{iau307}
\bibliography{MyBiblio}

\end{document}